\documentclass[conference]{IEEEtran}
\usepackage{cite}
\usepackage{amsmath,amssymb,amsfonts}
\usepackage[noend]{algpseudocode}
\usepackage{algorithm}
\usepackage{graphicx}
\usepackage{textcomp}
\usepackage{xcolor}
\usepackage[hyphens]{url}
\usepackage{mathtools}
\usepackage{threeparttable}
\usepackage{authblk}
\usepackage{hyperref}
\hypersetup{
    colorlinks=true,
    linkcolor=blue,
    filecolor=magenta,      
    urlcolor=blue,
    pdftitle={Overleaf Example},
    pdfpagemode=FullScreen,
    }

\DeclarePairedDelimiter{\ceil}{\lceil}{\rceil}
\DeclarePairedDelimiter{\floor}{\lfloor}{\rfloor}
\newcommand{\calO}{\mathcal{O}}

\def\BibTeX{{\rm B\kern-.05em{\sc i\kern-.025em b}\kern-.08em
    T\kern-.1667em\lower.7ex\hbox{E}\kern-.125emX}}

\title{A New Acceleration Paradigm for Discrete Cosine Transform and Other Fourier-Related Transforms}

\author[1]{Zixuan Jiang}
\author[1]{Jiaqi Gu}
\author[1]{David Z. Pan}
\affil[1]{Electrical and Computer Engineering, The University of Texas at Austin}
\affil[ ]{\textit{ \{zixuan, jqgu\}@utexas.edu, dpan@ece.utexas.edu}}

\begin{document}
\maketitle
\thispagestyle{plain}
\pagestyle{plain}

\begin{abstract}
Discrete cosine transform (DCT) and other Fourier-related transforms have broad applications in scientific computing.
However, off-the-shelf high-performance multi-dimensional DCT (MD DCT) libraries are not readily available in parallel computing systems.
Public MD DCT implementations leverage a straightforward method that decomposes the computation into multiple 1D DCTs along every single dimension,
which inevitably has non-optimal performance due to low computational efficiency, parallelism, and locality.
In this paper, we propose a new acceleration paradigm for MD DCT.
A three-stage procedure is proposed to factorize MD DCT into MD FFT and highly-optimized preprocessing/postprocessing with efficient computation and high arithmetic intensity.
Our paradigm can be easily extended to other Fourier-related transforms and other parallel computing systems.
Experimental results show that our 2D DCT/IDCT CUDA implementation has a stable, FFT-comparable execution time, which is $2\times$ faster than the previous row-column method.
Several case studies demonstrate that a promising efficiency improvement can be achieved with our paradigm.
The implementations are available at this \href{https://github.com/JeremieMelo/dct_cuda/tree/reconstruct}{link}.
\end{abstract}

\section{Introduction}
\label{sec:Introduction}

\begin{figure*}[htb]
    \centering
    \vspace{-5pt}
    \includegraphics[width=\textwidth]{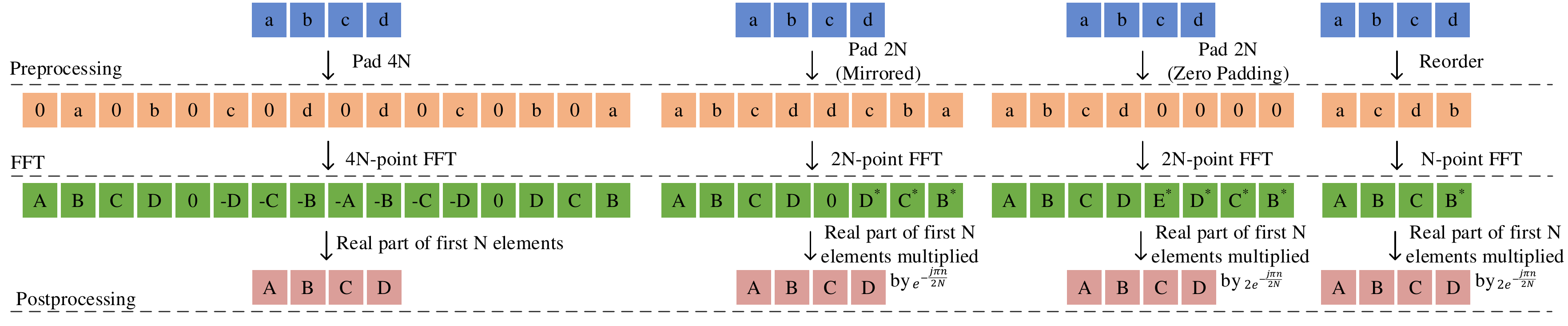}
    \caption{Four algorithms of 1D DCT using 1D FFT.
             All of them involve three steps, preprocessing, real to complex FFT, and postprocessing.
             Capital letters stand for complex numbers, while lower case letters represent real numbers.}
    \label{figure:1d-dct}
\end{figure*}

Discrete cosine transform (DCT) and other Fourier-related transforms provide frequency-domain information of the input signals, where different components oscillate at different frequencies.
They are of vital importance to various applications in science and engineering~\cite{dct-applications}.
We can use DCT/IDCT to perform digital signal processing~\cite{dctimageprocessing, dctimagedetection, nddct}, e.g., image/video compression~\cite{datacompression, dct3dcompression}.
DCT also plays a critical role in spectral methods for solving partial differential equations since most spectral methods require frequent function calls of DCT \cite{MEHRA201283}.

Given the performance limitation of sequential computing platforms, more computing tasks are off-loaded to parallel computing platforms, e.g., GPUs, multi-threaded CPUs.
Such parallel computing architectures are capable of achieving order-of-magnitude acceleration since many threads can be launched for concurrent computation. 
Therefore, the high-performance implementation of DCT and other Fourier-related transforms is necessary to fully leverage the massive parallelism of modern processors.
Related scientific computing applications can be further boosted.

Extensive public libraries and implementations focus on parallel Fourier-related transforms using \textit{CUDA}~\cite{cuda},
a general purpose parallel computing platform and programming model for GPUs.
The built-in \textit{cuFFT} library ~\cite{TOOL_cufft} in CUDA is specially designed and highly optimized to provide high-performance fast Fourier transform (FFT) on GPUs.

However, off-the-shelf highly-optimized DCT implementations are currently lacking in CUDA libraries, especially for multi-dimensional DCT (MD DCT).
Most open-source CUDA DCT implementations~\cite{TOOL_dct_tf, TOOL_dct_8by8} focus on JPEG/MPEG compression by using grid-like DCT with $8 \times 8$ blocks~\cite{Frid2013AccelerationOD,dct3dcompression, jpeg-1}.
This tiling-style DCT implementation has specific applications and is not general enough to handle other applications.
For general MD DCT, previous implementations leverage a straightforward method that simply decomposes the computation into multiple calls of 1D DCT along every single dimension~\cite{ghetia2013implementation}.
This row-column method inevitably has non-optimal performance since it is non-optimized in three aspects.
First, although enough parallelism has been explored within each dimension, the row-column method suffers from low parallelism across multiple dimensions.
Second, redundant memory access between consecutive calls leads to nontrivial memory transaction overhead, where the data locality is not well exploited.
Third, there exists redundant computation in multiple 1D DCT calls as the symmetry characteristic and potential data reuse are not considered.

In this paper, we propose a new acceleration paradigm to provide a high-performance operator for MD Fourier-related transforms.
It is specially optimized for parallelism, locality, and computational efficiency.
We factorize these transforms to the three-stage procedure, i.e., \texttt{preprocessing}, \texttt{MD FFT}, and \texttt{postprocessing}.
Thus, we can leverage the highly-optimized MD FFT libraries to accelerate the computation.
We provide an example of 2D DCT with CUDA to demonstrate the paradigm.
With highly optimized preprocessing and postprocessing kernels, our method runs much faster than the widely-used row-column method.
Experimental results show that our CUDA 2D DCT is $2\times$ faster than the previous implementations.
Our paradigm can be easily applied to other transforms and parallel systems.
Also, our method has a stable runtime, which is insensitive to transform types.
Several case studies in parallel scientific computing demonstrate a promising efficiency improvement with our proposed paradigm.

Our contributions are highlighted as follows.
\footnote{The implementations are available at \url{https://github.com/JeremieMelo/dct_cuda/tree/reconstruct}.}
\begin{itemize}
    \item We propose a computation- and memory-efficient three-stage procedure to cast parallel MD DCT into MD real-valued N-point FFT and highly-optimized preprocessing and postprocessing steps.
          Our parallel 2D DCT outperforms previous methods with better parallelism, locality, and computation efficiency,
          which achieves $2\times$ speedup than the row-column method.
    \item Our proposed acceleration paradigm can be easily extended to other Fourier-related transforms and parallel computing platforms with stable, FFT-comparable execution time.
    \item Several case studies demonstrate that our method can further boost the performance of related scientific computing applications.
\end{itemize}

\section{Background: Discrete cosine transforms}
\label{sec:background}
In this section, we give a brief introduction to the definition and computation algorithms of DCT.

\subsection{Definitions}
Formally, the DCT is a linear, invertible function $f: \mathbf{R}^n \to \mathbf{R}^n$~\cite{dct}.
There are eight standard variants, of which four are commonly used.
The DCT (DCT type 2) and the inverse DCT (DCT type 3, or the IDCT) for one-dimensional length-$N$ sequence $x$ is shown in Equation~\eqref{1d-definition},
\begin{subequations}
\label{1d-definition}
\begin{align}
    \textrm{DCT: } X_k & = \sum_{n=0}^{N-1} x_n \cos{\big(\frac{\pi}{N}(n+\frac{1}{2})k\big)}, \\
    \textrm{IDCT: } X_k & = \frac{1}{2}x_0 + \sum_{n=1}^{N-1} x_n \cos{\big(\frac{\pi}{N}n(k + \frac{1}{2})\big)},
\end{align}
\end{subequations}
where $k = 0, 1, \dots, N-1$. 

Multidimensional DCT follows straightforwardly from the one-dimensional definitions.
They are simply a composition of DCTs along every single dimension.
For example, a two-dimensional DCT (2D DCT) of a matrix is simply the 1D DCT performed along the rows and then along the columns (or in a reversed order).
The 2D DCT is formulated in Equation~\eqref{2d-dct-definition},
\begin{equation}
\footnotesize
\begin{split}
    \label{2d-dct-definition}
    X_{k_1, k_2} = \sum_{n_1=0}^{N_1-1} \sum_{n_2=0}^{N_2-1} x_{n_1, n_2} \cos{\big(\frac{\pi}{N_1}(n_1+\frac{1}{2})k_1\big)} \cos{\big( \frac{\pi}{N_2}(n_2+\frac{1}{2})k_2 \big)}
\end{split}
\end{equation}
where $k_i = 0, 1, \dots, N_i - 1$.

\begin{figure}[h]
    \centering
    \includegraphics[width=0.46\textwidth]{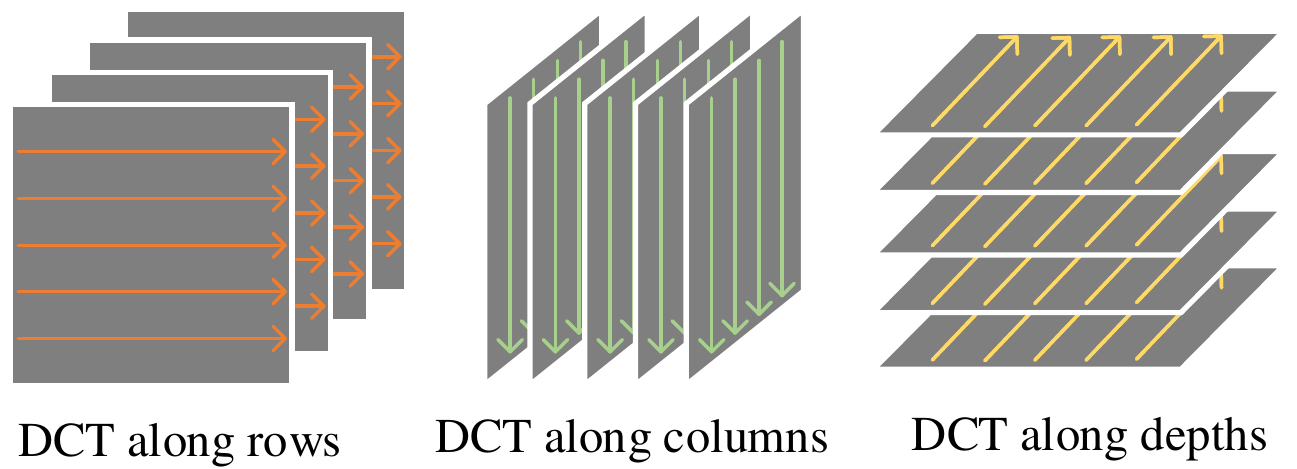}
    \caption{3D DCT follows straightforwardly from the definition of 1D DCT. 3D DCT are simply a separable DCTs along each dimension.}
    \label{figure:3d-dct}
\end{figure}

Figure~\ref{figure:3d-dct} illustrates a straightforward implementation of 3D DCT, where the 1D DCT is performed along the dimension of rows, columns, and depths, respectively.
Any order of the DCT along these three dimensions is valid.

\subsection{Computation of DCTs}

Although the definition of the DCT requires $\calO(N^2)$ operations, 
there are specialized algorithms to perform DCT, with $\calO(N\log N)$ complexity by factorizing the computation similarly to FFT.
We can also compute DCT via FFT combined with $\calO(N)$ preprocessing and postprocessing procedures.

Although DCT algorithms using FFT have the overhead of preprocessing and postprocessing compared to the specialized DCT algorithms,
they have a distinct advantage that FFT libraries are widely available and highly optimized.
The implementation and optimization, which requires substantial engineering insights and effort, significantly impact the ultimate performance.
Thus, we can usually obtain high performance for general length with FFT-based algorithms in practice.

Specialized DCT algorithms are usually used in the fixed sizes, such as the $8 \times 8$ DCT used in JPEG compression~\cite{TOOL_dct_8by8}.
The fixed size of DCT computation is not the focus of this paper.

\begin{algorithm}[htbp]
\begin{algorithmic}[1]
\caption{Four algorithms of 1D DCT via 1D FFT}
\label{1d-dct-via-fft}
\Require A real sequence $x$ with length of $N$ \Comment{$N$ can be any positive integer}

\Function{DCT using 4N FFT}{}
\State $x' = \texttt{dct\_4n\_preprocess}(x)$ using Eq.~\eqref{4n_pre}, where the length of $x'$ is $4N$
\begin{equation}
\label{4n_pre}
    x'(n) = 
        \begin{cases}
            x(\frac{n - 1}{2}),      & 0 \leqslant n < 2N, \text{$n$ is odd}, \\
            x(\frac{4N - n - 1}{2}), & 2N \leqslant n < 4N, \text{$n$ is odd}, \\
            0,                       & \text{$n$ is even}
        \end{cases}
\end{equation}
\State $X = \texttt{1D\_RFFT}(x')$; \Comment{1D real FFT kernel}
\State \Return $y = \texttt{dct\_4n\_postprocess}(X)$ using Eq.~\eqref{4n_post}, where the length of $y$ is $N$
\begin{equation}
\label{4n_post}
    y(n) = \operatorname{Re}(X(n))
\end{equation}
\EndFunction

\Function{DCT using mirrored 2N FFT}{}
\State $x' = \texttt{dct\_2n\_mirrored\_preprocess}(x)$ using Eq.~\eqref{2n_mirrored_pre}, where the length of $x'$ is $2N$
\begin{equation}
\label{2n_mirrored_pre}
    x'(n) = 
        \begin{cases}
            x(n),          & 0 \leqslant n < N \\
            x(2N - n - 1), & N \leqslant n < 2N
        \end{cases}
\end{equation}
\State $X = \texttt{1D\_RFFT}(x')$; \Comment{1D real FFT kernel}
\State \Return $y = \texttt{dct\_2n\_mirrored\_postprocess}(X)$ using Eq.~\eqref{2n_mirrored_post}, where the length of $y$ is $N$
\begin{equation}
\label{2n_mirrored_post}
    y(n) = \operatorname{Re}(e^{-\frac{j \pi n}{2N}} X(n))
\end{equation}
\EndFunction

\Function{DCT using padded 2N FFT}{}
\State $x' = \texttt{dct\_2n\_padded\_preprocess}(x)$ using Eq.~\eqref{2n_padded_pre}, where the length of $x'$ is $2N$
\begin{equation}
\label{2n_padded_pre}
    x'(n) = 
        \begin{cases}
            x(n), & 0 \leqslant n < N \\
            0,    & N \leqslant n < 2N
        \end{cases}
\end{equation}
\State $X = \texttt{1D\_RFFT}(x')$; \Comment{1D real FFT kernel}
\State \Return $y = \texttt{dct\_2n\_padded\_postprocess}(X)$ using Eq.~\eqref{2n_padded_post}, where the length of $y$ is $N$
\begin{equation}
\label{2n_padded_post}
    y(n) = 2\operatorname{Re}(e^{-\frac{j \pi n}{2N}} X(n))
\end{equation}
\EndFunction

\Function{DCT using N FFT}{}
\State $x' = \texttt{dct\_n\_preprocess}(x)$ using Eq.~\eqref{n_pre}, where the length of $x'$ is $N$
\begin{equation}
\label{n_pre}
    x'(n) = 
        \begin{cases}
            x(2n),            & 0 \leqslant n \leqslant \floor{\frac{N - 1}{2}} \\
            x(2N -2n - 1),    & \floor{\frac{N + 1}{2}} \leqslant n < N
        \end{cases}
\end{equation}
\State $X = \texttt{1D\_RFFT}(x')$; \Comment{1D real FFT kernel}
\State \Return $y = \texttt{dct\_n\_postprocess}(X)$ using Eq.~\eqref{n_post} or Eq.~\eqref{n_post_update}, where the length of $y$ is $N$
\begin{equation}
\label{n_post}
    y(n) = 2\operatorname{Re}(e^{-\frac{j \pi n}{2N}} X(n)), 0 \leqslant n < N
\end{equation}
\begin{equation}
\label{n_post_update}
    y(n) = 
        \begin{cases}
            2\operatorname{Re}(e^{-\frac{j \pi n}{2N}} X(n)),           &  0 \leqslant n \leqslant \ceil{\frac{N - 1}{2}} \\
            2\operatorname{Re}(e^{-\frac{j \pi n}{2N}} \overline{X(N - n)}), & \ceil{\frac{N + 1}{2}} \leqslant n < N
        \end{cases}
\end{equation}
\EndFunction
\end{algorithmic}
\end{algorithm}
\subsection{1D DCT through 1D FFT}

DCT can be performed via $\calO(N)$ sequence reordering, FFT, and $\calO(N)$ postprocessing steps.
There are four commonly used algorithms in this approach \cite{TOOL_dct_fft}, \cite{makhoul}. 
Figure~\ref{figure:1d-dct} and Algorithm~\ref{1d-dct-via-fft} demonstrate four algorithms of 1D DCT computations using 1D FFT, denoted 4N-point DCT, mirrored 2N-point DCT, padded 2N-point DCT~\cite{tensorflow} and N-point DCT, respectively.
The N-point algorithm is the fastest among these four methods since the preprocessing, FFT, and postprocessing only need to handle the sequence of length $N$.
Therefore, we focus on the N-point algorithm in the following discussions.
Similar principles apply to the other three algorithms.

Since the DCT is performed on a real-valued tensor, the elements of FFT input are all real numbers.
For real-valued input sequence of length $N$, the 1D FFT outputs have the conjugate symmetry (also known as the Hermitian symmetry)
\begin{equation}
    \label{equation:conjugate-symmetry}
    X(n) =X^{\ast}(N - n)
\end{equation}
where $\ast$ is the conjugate operator.
~\footnote{The complex conjugate of a complex number $x$ can also be written as $\overline{x}$.
Its real part and imaginary part are $\operatorname{Re}(x)$ and $\operatorname{Im}(x)$, respectively.
We use $j$ to represent the imaginary unit ($j^2 = -1$) in this paper.}
Note that we use zero-based array indexing throughout the paper. Therefore, we have $1 \leq n \leq N-1$ in Equation~\eqref{equation:conjugate-symmetry}.
~\footnote{
We provide two examples of the conjugate symmetry in Equation~\eqref{equation:conjugate-symmetry}.
If $N=5$, then $X(1) =X^{\ast}(4), X(2) =X^{\ast}(3)$.
If $N=6$, then $X(1) =X^{\ast}(5), X(2) =X^{\ast}(4), X(3) =X^{\ast}(3)$.}

Efficient algorithms have been designed for the real-valued FFT (RFFT)~\cite{rfft}.
This characteristic is also considered in most FFT libraries.
For instance, cuFFT~\cite{TOOL_cufft} takes advantage of this redundancy and works only on the first half of the Hermitian vector.
Specifically, the output sequence length is $\ceil{\frac{N + 1}{2}}$.
The postprocessing in Algorithm~\ref{1d-dct-via-fft} can be changed to leverage the Hermitian symmetry.
Precisely, Equation~\eqref{n_post} should be replaced with Equation~\eqref{n_post_update}.

\subsection{2D DCT using 2D FFT}
\begin{algorithm*}[h]
\caption{2D DCT, 2D IDCT, with $N$-Point 2D FFT}
\label{alg:DCT}
\setlength{\columnwidth}{\linewidth}
\begin{algorithmic}[1]
\Require An real $N_1 \times N_2$ matrix $x$;  \Comment{$N_1$ and $N_2$ can be any positive integer}

\Function{2D\_DCT}{$x$}
\State $x' = \texttt{2d\_dct\_preprocess}(x)$ using Eq.~\eqref{2d_dct_pre},
\begin{equation}
\label{2d_dct_pre}
  x'(n_1, n_2) =
    \begin{cases}
      x(2n_1, 2n_2),                 & 0 \leqslant n_1 \leqslant \floor*{\frac{N_1 - 1}{2}}, 0 \leqslant n_2 \leqslant \floor*{\frac{N_2 - 1}{2}}, \\
      x(2N_1 - 2n_1 -1, 2n_2),       & \floor*{\frac{N_1+1}{2}} \leqslant n_1 < N_1, 0 \leqslant n_2 \leqslant \floor*{\frac{N_2 - 1}{2}}, \\
      x(2n_1, 2N_2-2n_2-1),          & 0 \leqslant n_1 \leqslant \floor*{\frac{N_1 - 1}{2}}, \floor*{\frac{N_2+1}{2}} \leqslant n_2 < N_2, \\
      x(2N_1 - 2n_1-1, 2N_2-2n_2-1), & \floor*{\frac{N_1+1}{2}} \leqslant n_1 < N_1, \floor*{\frac{N_2+1}{2}} \leqslant n_2 < N_2
    \end{cases}
\end{equation}
\State $X = \texttt{2D\_RFFT}(x')$; \Comment{2D real FFT kernel}
\State \Return $y = \texttt{2d\_dct\_postprocess}(X)$ using Eq.~\eqref{2d_dct_post},
\begin{equation}
\label{2d_dct_post}
\begin{split}
      y(n_1, n_2) & = 2 \operatorname{Re}\Big(e^{-\frac{j \pi n_2}{2N_2}} \big(e^{-\frac{j \pi n_1}{2N_1}} X(n_1, n_2) + e^{\frac{j \pi n_1}{2N_1}} X(N_1 - n_1, n_2)\big)\Big), \\
      \textrm{where } & X(N_1, n_2) = X(n_1, N_2) = 0, \forall n_1, n_2; 
\end{split}
\end{equation}

\EndFunction

\Function{2D\_IDCT}{$x$}
\State $x' = \texttt{2d\_idct\_preprocess}(x)$ using Eq.~\eqref{2d_idct_pre}
\begin{equation}
\label{2d_idct_pre}
\begin{split}
     x'(n_1, n_2) & = e^{-\frac{j \pi n_1}{2N_1}} e^{-\frac{j \pi n_2}{2N_2}} \Big(x(n_1, n_2) -x(N_1 - n_1, N_2-n_2) - j \big(x(N_1 -n_1, n_2) + x(n_1, N_2 - n_2)\big)\Big), \\
     \textrm{where } & x(N_1, n_2) = x(n_1, N_2) = 0, \forall n_1, n_2; 
\end{split}
\end{equation}
\State $X = \texttt{2D\_IRFFT}(x')$; \Comment{2D real inverse FFT kernel}
\State \Return $y = \texttt{2d\_idct\_postprocess}(X) = \texttt{2d\_dct\_preprocess}^{-1}(X)$ using Eq.~\eqref{2d_idct_post}
\begin{equation}
\label{2d_idct_post}
  y(n_1, n_2) =
    \begin{cases}
      X(\frac{n_1}{2}, \frac{n_2}{2}),                     & \text{$n_1$ is even, $n_2$ is even}, \\
      X(N_1 - \frac{n_1 + 1}{2}, \frac{n_2}{2}),           & \text{$n_1$ is odd, $n_2$ is even}, \\
      X(\frac{n_1}{2}, N_2 - \frac{n_2 + 1}{2}),           & \text{$n_1$ is even, $n_2$ is odd}, \\
      X(N_1 - \frac{n_1 + 1}{2}, N_2 - \frac{n_2 + 1}{2}), & \text{$n_1$ is odd, $n_2$ is odd}
    \end{cases}
\end{equation}
\EndFunction

\end{algorithmic}
\end{algorithm*}

In previous implementations, 2D DCT is calculated by first performing 1D DCT through rows and 1D DCT through columns. 
Nevertheless, it is limited by the sequential nature of two 1D DCT and expensive redundant memory access.
The overhead of function-call in invoking FFT routines and the redundant global memory access are quite significant on GPUs.
We can achieve higher performance by reducing function calls and memory access, which requires us to manipulate the matrix in a 2D fashion instead of a row-column way.

Similar to the algorithm of 1D DCT using 1D FFT, 2D DCT can also be conducted via 2D FFT of the same input size~\cite{makhoul}, as shown in Algorithm~\ref{alg:DCT}.
In 2D DCT, the preprocessing reorder input sequence such that odd or even sub-sequences are grouped in a grid fashion, described in Line 2.
It is equivalent to the 1D reordering along rows and columns, respectively.
Then 2D real-valued FFT is applied in Line 3. 
The final step is the postprocessing, which combines the FFT results, scales each element by a particular exponential coefficient, and extracts the real part as the final result. 
Since we use zero-based array indexing throughout the paper, $\forall n_1, n_2, X(N_1, n_2) \text{ and } X(n_1, N_2)$ are not defined originally.
We define it as $0$ in Equation~\eqref{2d_dct_post} to simplify the algorithm description.

For 2D IDCT, preprocessing and postprocessing are similar to the 2D DCT counterpart but are switched according to the mathematical formula.

For real-valued input data, the 2D FFT outputs have the conjugate symmetry $X(n_1, n_2) =X^{\ast}(N_1 - n_1, N_2 - n_2)$.
Similarly, for IFFT with real number input data, only half of the input is needed.
We will leverage this symmetry to optimize the postprocessing of 2D DCT and the preprocessing of 2D IDCT in Section~\ref{sec:implementation-post}.

Following the convention of computation complexity and time complexity in parallel algorithms, 
we list the work and depth
\footnote{
The \textit{work} of a parallel algorithm is the total number of primitive operations that the processors perform~\cite{casanova2008parallel}.
It is the time used to run the computation on a single processor.
The \textit{depth} or \textit{span} is the length of the longest series of operations that have to be performed sequentially due to data dependencies.
The depth can also be defined as the time using an idealized machine with an infinite number of processors.}
for each step in Table~\ref{tab:WorkDepth2DDCT}. 
Our implementations are \textit{work optimal} since we share the same work complexity with the best sequential algorithm.

\begin{table}[h]
\caption{Work and depth analysis of 2D DCT via 2D FFT}
\centering
\label{tab:WorkDepth2DDCT}
\begin{tabular}{ccc}
\hline \hline
                & Work & Depth \\ \hline
Preprocessing  & $\calO(N_1N_2)$    & $\calO(1)$     \\
2D FFT          & $\calO(N_1N_2\log{N_1N_2})$   & $\calO(\log{N_1N_2})$     \\
Postprocessing & $\calO(N_1N_2)$    & $\calO(1)$     \\
Total           & $\calO(N_1N_2\log{N_1N_2})$ & $\calO(\log{N_1N_2})$ \\
\hline \hline
\end{tabular}
\end{table}

\section{Proposed Methodology and Algorithms}
\label{sec:Methodology}

We propose to cast DCT computation into three stages: preprocessing, real FFT (RFFT), and postprocessing.
This section will take the 2D DCT with CUDA as an example to demonstrate the performance benefits from our proposed paradigm.
Specifically, we demonstrate how we accelerate Lines 2 and 4 of Algorithm~\ref{alg:DCT} in Sections~\ref{sec:efficient-preprocessing} and~\ref{sec:implementation-post}, respectively.

\subsection{Efficient Preprocessing}
\label{sec:efficient-preprocessing}
\begin{figure}[h]
    \centering
    \includegraphics[width=0.46\textwidth]{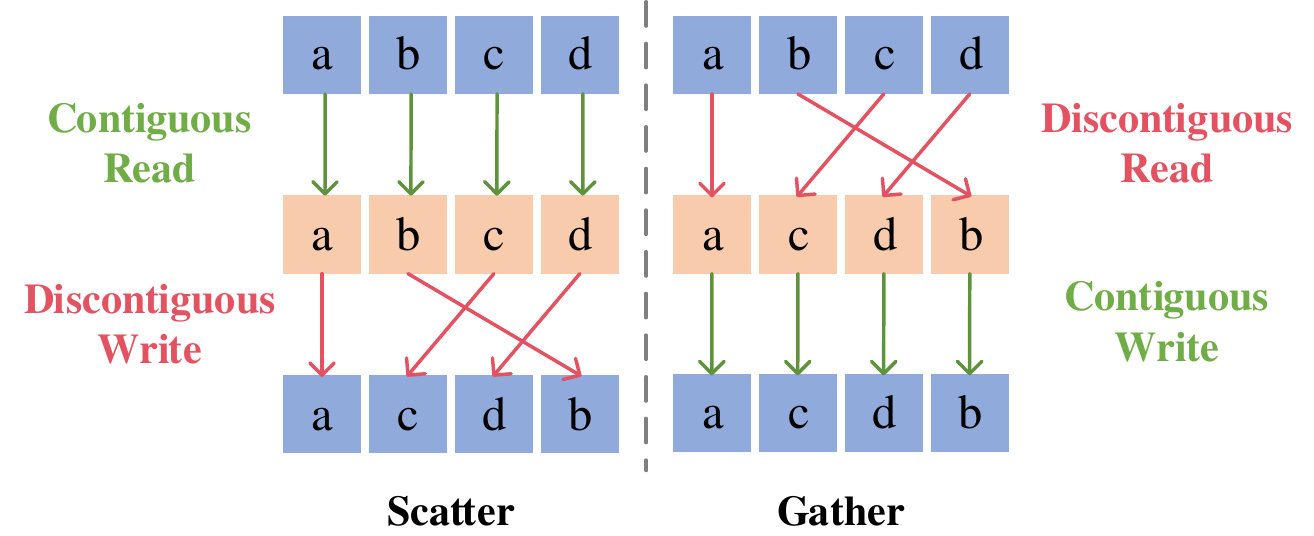}
    \caption{Two routines of reordering memory.}
    \label{fig:memory_reorder}
\end{figure}

In the preprocessing of the 2D DCT and the postprocessing of the 2D IDCT, we reorder a matrix of shape $N_1\times N_2$.
The CUDA kernel launches $N_1\times N_2$ threads to perform fully parallel reordering.
Each CUDA thread handles one element of the matrix, involving reading from the source and writing to the destination.
The order of elements processed has a significant impact on the performance of the CUDA kernels.
If 32 threads in a GPU warp access global memory in arbitrary discontiguous addresses that are misaligned with 128 bytes for L1 cache blocks or 32 bytes for L2 cache, 
redundant memory access will occur, leading to a considerable runtime penalty.
Therefore, following specific memory access patterns will better utilize the global memory bandwidth through \textit{memory coalescing} that merges multiple accesses by a single load/write instruction.
There are two routines to handle data reordering, \textit{gather} and \textit{scatter}, as shown in Figure~\ref{fig:memory_reorder}.

In the gather method, we can assign each thread to an output matrix element.
Specifically, all CUDA threads within a warp will read the element from the source matrix and write it to a consecutive memory in the output matrix.
This method directly leverages Equations~\eqref{2d_dct_pre} and~\eqref{2d_idct_post}, and it can achieve coalesced memory write at the cost of discontiguous memory read.

On the contrary, we can traverse the element in terms of the input matrix.
For every single element in the input matrix, we locate it in the output matrix and put it there.
Threads in the warp will read from a contiguous block in the source matrix and write the corresponding values to a discontiguous memory location. 
In this scatter method, a coalesced memory read is achieved at the expense of discontiguous memory write.

Scatter and gather are well-optimized operations, and coalesced read and write generally have similar performance in NVIDIA GPUs~\cite{gpuscattergather}, as shown in Table~\ref{tab:GatherScatterRuntime}.
In this paper, we perform tensor reordering using the scatter method, with each single thread processing one element of the tensor.
\begin{table}[h]
\caption{2D DCT preprocessing time (ms) with gather/scatter}
\label{tab:GatherScatterRuntime}
\centering
\begin{tabular}{cccccc}
\hline\hline
$N$       & 512   & 1024  & 2048  & 4096  & 8192  \\\hline
Gather  & 0.013 & 0.042 & 0.160 & 0.627 & 2.568 \\
Scatter & 0.014 & 0.043 & 0.163 & 0.633 & 2.524\\
\hline\hline
\end{tabular}
\end{table}

\subsection{Efficient Postprocessing}
\label{sec:implementation-post}

\begin{table*}[htbp]
\caption{Comparison of the naive method and our efficient method on the 2D DCT postprocessing}
\label{table:efficient-processing}
\centering
\begin{tabular}{cc|cccc|ccc}
\hline \hline
             &                     & \multicolumn{4}{c|}{per thread}                                 & \multicolumn{3}{c}{Total}                           \\
             & \# thread          & \# read & \# multiplication & \# addition & Arithmetic Intensity & \# read            & \# multiplication & \# addition \\ \hline
Naive method & $N_1N_2$    & 2       & 10              & 7            & 8.5                  & $2N_1N_2$          & $10N_1N_2$      & $7N_1N_2$    \\
Our method   & $N_1 N_2 / 4$ & 2       & 16              & 12           & 14                   & $N_1N_2 / 2$ & $4N_1N_2$       & $3N_1N_2$    \\ \hline \hline
\end{tabular}
\begin{tablenotes}
\small
\item
    We assume that $N_1$ and $N_2$ are both even in this table.
    The multiplications and additions are real-valued.
    Arithmetic intensity is defined as the number of computations per memory access, higher of which indicates better utilization of computation capacity,
    thus more likely to reach a better design point according to the roofline model~\cite{roofline}.
\end{tablenotes}
\end{table*}

In the postprocessing of the 2D DCT and the preprocessing of the 2D IDCT, 
we need to 
(1) read multiple elements of the input matrix,
(2) complete the shift computation,
and (3) write results to the corresponding positions.

The naive method of the postprocessing is to launch $N_1 \times N_2 $ threads.
The thread with index $(n_1, n_2)$ will compute the corresponding $y(n_1, n_2)$ following Equation~\eqref{2d_dct_post}.

We propose two techniques to improve the efficiency of this naive implementation.
First of all, we notice that in Equation~\eqref{2d_dct_post}, to obtain the results of $y(n_1, n_2)$ and $y(N_1 - n_1, n_2)$, we need to read the same elements of $X(n_1, n_2)$ and $X(N_1 - n_1, n_2)$.
Therefore, we can merge these two threads with indices of $(n_1, n_2)$ and $(N_1 - n_1, n_2)$ to avoid redundant memory access.
Second, due to the conjugate symmetry, $X(n_1, n_2) = X^{\ast}(N_1 - n_1, N_2 - n_2)$, the real-valued FFT can return half of results to avoid redundancy.
In other words, the shape of the output matrix $X$ is $(N_1, \ceil{\frac{N_2+1}{2}})$ instead of $(N_1, N_2)$.
When we read $X(n_1, n_2), n_2 > \ceil{N_2 / 2}$, we actually read $X(N_1 - n_1, N_2 - n_2)$ and calculate its complex conjugate.
The conjugate symmetry allows us to further improve the efficiency of the postprocessing.

Specifically, we can only invoke $\ceil{N_1 / 2} \times \ceil{N_2 / 2}$ threads in this procedure.
Each thread will undertake the computation tasks of four threads in the naive method.
For thread with the index $(n_1, n_2)$ where $0 < n_1 < \ceil{N_1 / 2}, 0 < n_2 < \ceil{N_2 / 2}$,
it will read two elements $X(n_1, n_2), X(N_1 - n_1, n_2)$ and compute four elements
\begin{subequations}
\label{eq:four-elements}
\begin{align}
    y(n_1, n_2) & = 2 \operatorname{Re}(s), \\
    y(N_1 - n_1, n_2) & = -2 \operatorname{Im}(t), \\
    y(n_1, N_2 - n_2) & = -2 \operatorname{Im}(s), \\
    y(N_1 - n_1, N_2 - n_2) & = -2 \operatorname{Re}(t)
\end{align}
\end{subequations}
where
\begin{subequations}
\label{eq:s-t}
\begin{align}
s & = b \big(a X(n_1, n_2) + \overline{a} X(N_1 - n_1, n_2)\big), \\
t & = b \big(a X(n_1, n_2) - \overline{a} X(N_1 - n_1, n_2)\big), \\
a & = e^{-\frac{j \pi n_1}{2N_1}}, \overline{a} = e^{\frac{j \pi n_1}{2N_1}}, b = e^{-\frac{j \pi n_2}{2N_2}}, \label{precompute}
\end{align}
\end{subequations}

The terms of $a$ and $b$ in Equations~\eqref{precompute} are pre-computed and fixed before the call of the DCT procedures, which is a standard convention to improve the efficiency in repeated function calls.
We leverage the conjugate symmetry between $a$ and $\overline{a}$ to avoid storing $\overline{a}$.


The thread with the index of $(n_1, n_2), n_1 = 0 \text{ or } n_2 = 0$ is the corner case.
We follow Equation~\eqref{2d_dct_post} and assign the same workload to these threads.
In other words, every single thread will process four elements in the output matrix, which induces the workload balance across threads.
\footnote{If $N_1$ or $N_2$ is odd, partial threads will process 2 elements instead.}
Our method ensures that each element of the input/output matrix will be read/written only once.
In other words, there is no overlap between different threads in terms of memory access and computation.

Table~\ref{table:efficient-processing} compares the naive method and our efficient postprocessing method.
The straightforward implementation launches one thread for each output element and follows Equation~\eqref{2d_dct_post}.
In our method, each element in the input matrix will be read only once, and there are no redundant computations since we have fully leveraged the symmetry.
Compared with the straightforward approach, our method has a better arithmetic intensity.
The number of memory reads and the number of computations decrease significantly.

\begin{figure*}[h]
    \centering
    \includegraphics[width=0.8\textwidth]{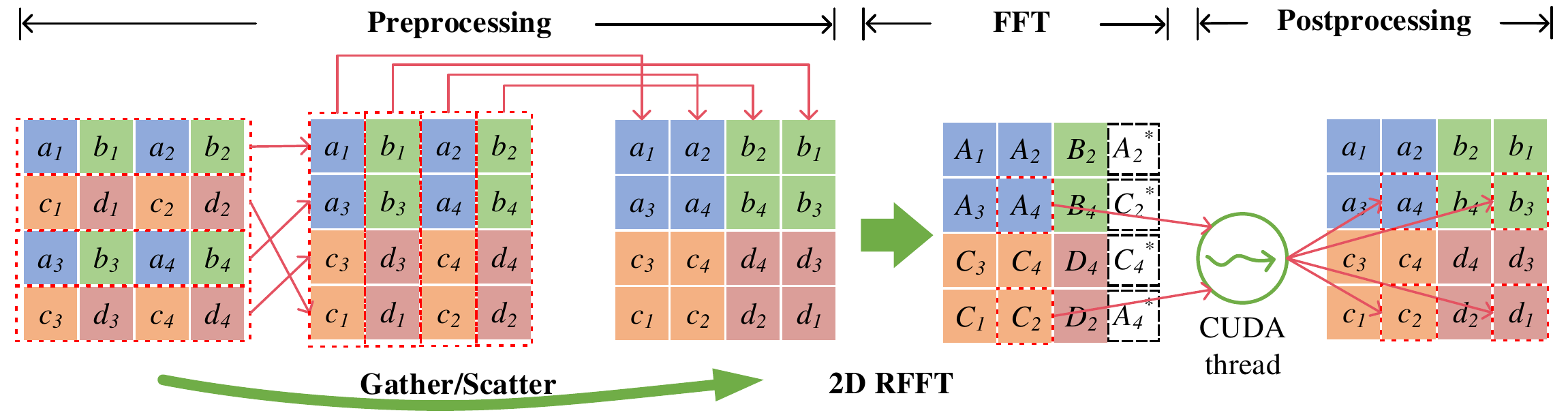}
    \caption{An example of the 2D DCT in our implementations. Capital and lower letters represent complex and real numbers, respectively.}
    \label{fig:2d-example}
\end{figure*}

Similarly, in the preprocessing of the 2D IDCT, each thread reads four elements from the input matrix and writes two elements to the output matrix.
We fully exploit the conjugate symmetry for higher efficiency.

\subsection{Examples and Summary}

Figure~\ref{fig:2d-example} illustrates a 2D DCT with a $4 \times 4$ input matrix.
In the preprocessing stage, the input matrix's reordering can be depicted as two butterfly reorderings along the row and column, respectively.
For better memory access efficiency, we perform the reordering in one step for the 2D input.
Afterward, 2D RFFT is performed on the reordered input, and we obtain the onesided complex-valued results.
The last row will not be returned to avoid redundancy.
Finally, in the postprocessing stage, each thread will read and process two complex numbers and finally write four real-valued results to the corresponding positions.

\begin{figure}[htbp]
    \centering
    \vspace{-15pt}
    \includegraphics[width=0.45\textwidth]{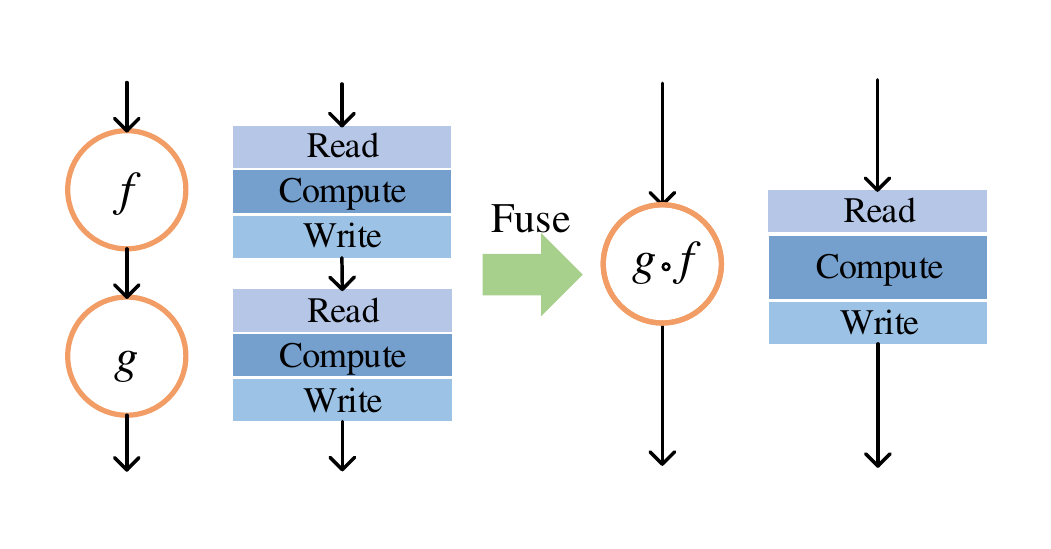}
    \caption{\textbf{Left:} Row-column methods need to call 1D transforms repeatedly.
             \textbf{Right:} MD DCT using MD FFT is equivalent to fusing the separate operators so that we can reduce memory transactions.
             Also, the operator composition $g \circ f$ provides us an opportunity to remove the redundant computations in functions $f$ and $g$.}
    \label{fig:operator-fusion}
\end{figure}

Compared with the row-column methods, the computation of 2D DCT using 2D FFT can be viewed as an operator fusion~\cite{operatorfusion}, as shown in Figure~\ref{fig:operator-fusion}.
In the row-column methods, we call the 1D DCT kernel twice with two necessary transpose operations.
Each 1D DCT kernel also involves a complete procedure of preprocessing, 1D RFFT, and postprocessing.
Thus, there are $3\times2+2=8$ stages of full-matrix global memory transactions, as well as redundant computations.

On the contrary, performing 2D DCT via 2D FFT involves only $3$ stages of memory access, which saves $\sim\!\!62.5\%$ memory access compared to its row-column counterpart.
Further, by fusing the functions, we can remove repeating operations in the row-column methods.
That is the reason why we can accelerate 2D DCT.

\subsection{Other Transforms and Platforms}
As long as the Fourier-related transforms can be computed via FFT with preprocessing and postprocessing, they can be accelerated using our paradigm.
We will show how we extend our framework to other transforms in Section~\ref{sec:case_studies}.

Our method can apply to multi-dimensional transforms.
Given that cuFFT only provides 1D, 2D, and 3D FFT APIs, we can use them directly to compute 1D, 2D, and 3D DCT.
Specifically, our method in 2D transforms can be naturally extended to 3D transforms with similar implementations.
The preprocessing and postprocessing of 3D DCT are similar to those of 2D DCT.
The preprocessing reorders the input 3D tensor with standard gather/scatter operations.
For the postprocessing, each thread reads 4 elements from the input tensor and writes 8 elements to the output tensor.

For higher dimension DCT, we can factorize them into lower dimensions, which is similar to how we factorize the MD DCT into the 1D DCT computations.
For example, a 4D DCT can be factorized into two rounds of 2D DCTs.
We can compute the DCT along any two dimensions at first and then perform DCT along the other two dimensions.

Our method can be easily transferred to other parallel systems if there exists an FFT library.
An example is to extend our method to multi-core CPUs. 
With similar optimization techniques in the preprocessing and postprocessing, we can use the highly-optimized FFT library on multi-core CPUs, e.g., FFTW~\cite{fftw}.

An interesting question is how our method scales to systems with multiple GPUs.
Our framework can well support computation with multiple GPUs for both batched MD DCTs and a single large MD DCT instance.
For batched MD DCTs, the task can be embarrassingly parallelized on multiple GPUs.
Hence, the speedup approximately scales to the number of GPUs.
For a single large-scale MD DCT, we can use the standard communication operations to distribute/collect data to/from all GPUs in preprocessing and postprocessing stages.
Note that our preprocessing and postprocessing stages have no data dependency or write conflicts across different threads.
Specifically, our method ensures that each element of the input/output tensor will be read/written only once. 
There is no overlap between different threads in terms of memory access and computation.
Hence, we can distribute these computation loads across GPUs easily without complicated memory patterns and computation dependency.
The MD RFFT can also be efficiently mapped to multiple GPUs by using the built-in cuFFT library.
\footnote{Following the cuFFT official user's guide~\cite{TOOL_cufft}, cuFFT supports using up to 16 GPUs connected to a CPU to perform FFTs whose calculations are distributed across the GPUs.
Multiple GPU execution is not guaranteed to solve a given size problem in a shorter time than single GPU execution.}


\section{Experiments}
\label{sec:experiments}

In this section, we conduct experiments and discuss the results of our implementations.

\subsection{Settings}
We implement the algorithms in C++/CUDA.
All the programs run on a Linux server with Intel Core i9-7900X @ 3.30GHz and 1 NVIDIA Titan Xp GPU. 
We use the RFFT and IRFFT kernels in cuFFT.

We first ensure that our implementation obtains correct results.
Unless stated otherwise, we report the average execution time of 100 runs and use the datatype of double-precision floating-point for performance evaluation.
The execution time is independent and identically distributed, and has a very low variance.
For example, the execution time of $1024 \times 1024$ 2D DCT is $3.765 \times 10^{-1}$ ms with a standard deviation of $3.196 \times 10^{-3}$ ms.
In all our experiments, we find that the standard deviation is less than $1\%$ of the average.
The speedup on the single-precision floating-point is similar to that of the double-precision floating-point.
For integer inputs, we have to cast them as the single-precision floating-point since the cuFFT does not support integer FFTs.

In real applications, the workloads of Fourier-related transforms are typically static.
The time for computing $\{e^{-\frac{j \pi n}{2N}}\}_{n=0}^{N-1}$ can be fully amortized by multiple procedure calls.
Therefore, we pre-compute the above coefficient sequence to prevent redundant computation and exclude its execution time in our evaluations.

\subsection{Four Algorithms of 1D DCT}
\label{sec:1dResults}
We first evaluate the performance of the four algorithms for 1D DCT, i.e., 4N-point DCT, mirrored 2N-point DCT, padded 2N-point DCT, and N-point DCT, respectively.
Similar to the 2D DCT computations discussed in Section~\ref{sec:Methodology}, we optimize the preprocessing and postprocessing of 1D computations.
We list the results in Table~\ref{table:1ddct}.
The N-point algorithm is the fastest one among these four methods,
since the preprocessing, FFT, and postprocessing only need to handle the sequence of length $N$.

\begin{table}[htbp]
\caption{Execution time in microseconds of four algorithms of 1D DCT via 1D FFT}
\label{table:1ddct}
\centering
\begin{tabular}{ccccc}
\hline \hline
Input size $N$ & {4N}      & {Mirrored 2N} & {Padded 2N} & {N}      \\ \hline
{$2^{14}$}                     & {190.41}  & {155.34}      & {144.32}    & {101.62} \\
{$2^{15}$}                      & {292.34}  & {207.29}      & {208.76}    & {122.60} \\
{$2^{16}$}                      & {416.20}  & {301.66}      & {309.32}    & {133.50} \\
{$2^{17}$}                      & {639.64}  & {414.04}      & {443.18}    & {158.96} \\
{$2^{18}$}                      & {1099.31} & {645.24}      & {652.23}    & {215.99} \\ \hline \hline
\end{tabular}
\end{table}


\subsection{Results on 2D DCTs}
\label{sec:2dResults}

\begin{table*}[h]
\caption{Execution time in milliseconds and the ratio to our 2D DCT/IDCT implementations}
\label{table:2dresult}
\centering
\begin{tabular}{cc|cccc|ccc}
\hline\hline
$N_1$ & $N_2$ & \begin{tabular}[c]{@{}c@{}}DCT2D\\ MATLAB\end{tabular} & \begin{tabular}[c]{@{}c@{}}DCT2D\\ row-column\end{tabular} & \textbf{\begin{tabular}[c]{@{}c@{}}DCT2D\\ via RFFT2D\end{tabular}} & RFFT2D       & \begin{tabular}[c]{@{}c@{}}IDCT2D\\ row-column\end{tabular} & \textbf{\begin{tabular}[c]{@{}c@{}}IDCT2D\\ via IRFFT2D\end{tabular}} & IRFFT2D      \\ \hline
512   & 512   & 2.43 (20.18)                                           & 0.19 (1.61)                                                                                         & \textbf{0.12 (1)}                                                   & 0.10 (0.81)  & 0.24 (1.87)                                                 & \textbf{0.13 (1)}                                                     & 0.10 (0.81)  \\
1024  & 1024  & 9.83 (26.34)                                           & 0.66 (1.76)                                                                                         & \textbf{0.37 (1)}                                                   & 0.33 (0.88)  & 0.88 (2.10)                                                 & \textbf{0.42 (1)}                                                     & 0.35 (0.83)  \\
2048  & 2048  & 30.62 (20.91)                                          & 2.58 (1.76)                                                                                         & \textbf{1.46 (1)}                                                   & 1.12 (0.77)  & 3.45 (2.13)                                                 & \textbf{1.62 (1)}                                                     & 1.28 (0.79)  \\
4096  & 4096  & 122.79 (21.15)                                         & 12.28 (2.11)                                                                                                & \textbf{5.80 (1)}                                                   & 4.44 (0.77)  & 15.80 (2.45)                                                & \textbf{6.44 (1)}                                                     & 5.04 (0.78)  \\
8192  & 8192  & 525.03 (20.37)                                         & 54.18 (2.10)                                                                                                & \textbf{25.78 (1)}                                                  & 20.11 (0.78) & 67.39 (2.35)                                                & \textbf{28.64 (1)}                                                    & 22.50 (0.79) \\ \hline
100   & 10000 & 9.52 (20.85)                                           & 1.04 (2.29)                                                                                                 & \textbf{0.46 (1)}                                                   & 0.39 (0.86)  & 1.44 (2.82)                                                 & \textbf{0.51 (1)}                                                     & 0.41 (0.81)  \\
10000 & 100   & 9.47 (20.74)                                           & 1.03 (2.26)                                                                                                 & \textbf{0.46 (1)}                                                   & 0.48 (1.05)  & 1.43 (2.80)                                                 & \textbf{0.51 (1)}                                                     & 0.49 (0.96)  \\ \hline\hline
\end{tabular}
\end{table*}

For 2D DCT, we compare our implementation with MATLAB implementations on GPUs, the traditional row-column method on GPUs.
For 2D IDCT, we compare our implementation with the traditional row-column method.
\footnote{The \texttt{idct2} API in MATLAB does not support GPU acceleration.}
We do not find public implementations on 2D DCT via 2D RFFT.
For the row-column methods, the public implementations~\footnote{An example is \url{https://github.com/zh217/torch-dct}} are less optimized.
We implement and optimize the row-column method based on our 1D DCT/IDCT implementation, which is better than the public implementations we can find.

We list all tested execution time (ms) in Table~\ref{table:2dresult}.
The execution time of 2D RFFT/IRFFT is also measured for reference.

Our optimized version could achieve more than $2 \times$ speedup than the row-column methods on different input sizes.
This speedup is quite stable across different input sizes.
As analyzed before, the total memory access and computations of our method are approximately reduced by $\sim$60\% due to fewer stages of memory transactions and less redundant computations.
Thus a roughly constant speedup around $2 \times$ compared with the row-column method can be observed on various DCT workloads.
Besides, the gap between our implementation and 2D RFFT indicates the overhead introduced by our preprocessing and postprocessing are very small,
which means 2D RFFT/IRFFT dominates the execution time.
As problem size increases, the execution time of the row-column method increases dramatically, while our implementations demonstrate superior scalability.

In the last two rows of Table~\ref{table:2dresult}, we observe that RFFT/IRFFT in cuFFT performs better at the cases where $N_2 \gg N_1$.
We can further leverage this feature in 2D DCT/IDCT.
In particular, given the input of the 2D DCT/IDCT with shape of $(N_1, N_2)$, we can transpose the input and output if $N_2 \gg N_1$.
Since the transpose operation can be fused with the preprocessing and postprocessing without overhead, the computation time on the input with shape $(N_1, N_2)$ is the same as the input with shape $(N_2, N_1)$.
Thus, the execution time of the DCT in the case of $10^4 \times 10^2$ is even better than that of the RFFT.

We also compare the performance with other public CUDA MD DCT implementations. 
TensorFlow2.1~\footnote{The link to the documentation is \url{https://www.tensorflow.org/api_docs/python/tf/signal/dct}. TensorFlow only provides the API of 1D DCT/IDCT. We implement the row-column method based on that.} and MATLAB2019a demonstrate $3 - 4\times$ and $\sim 20 \times$ slower than our implementations respectively.
To fully show our advantages, we implement our own optimized row-column methods, which are already 10x faster than MATLAB, as an even stronger baseline. Our claimed 2x speedup is more convincing since it is compared with the competitive row-column baseline.

\begin{figure}[t]
    \centering
    \includegraphics[width=0.25\textwidth]{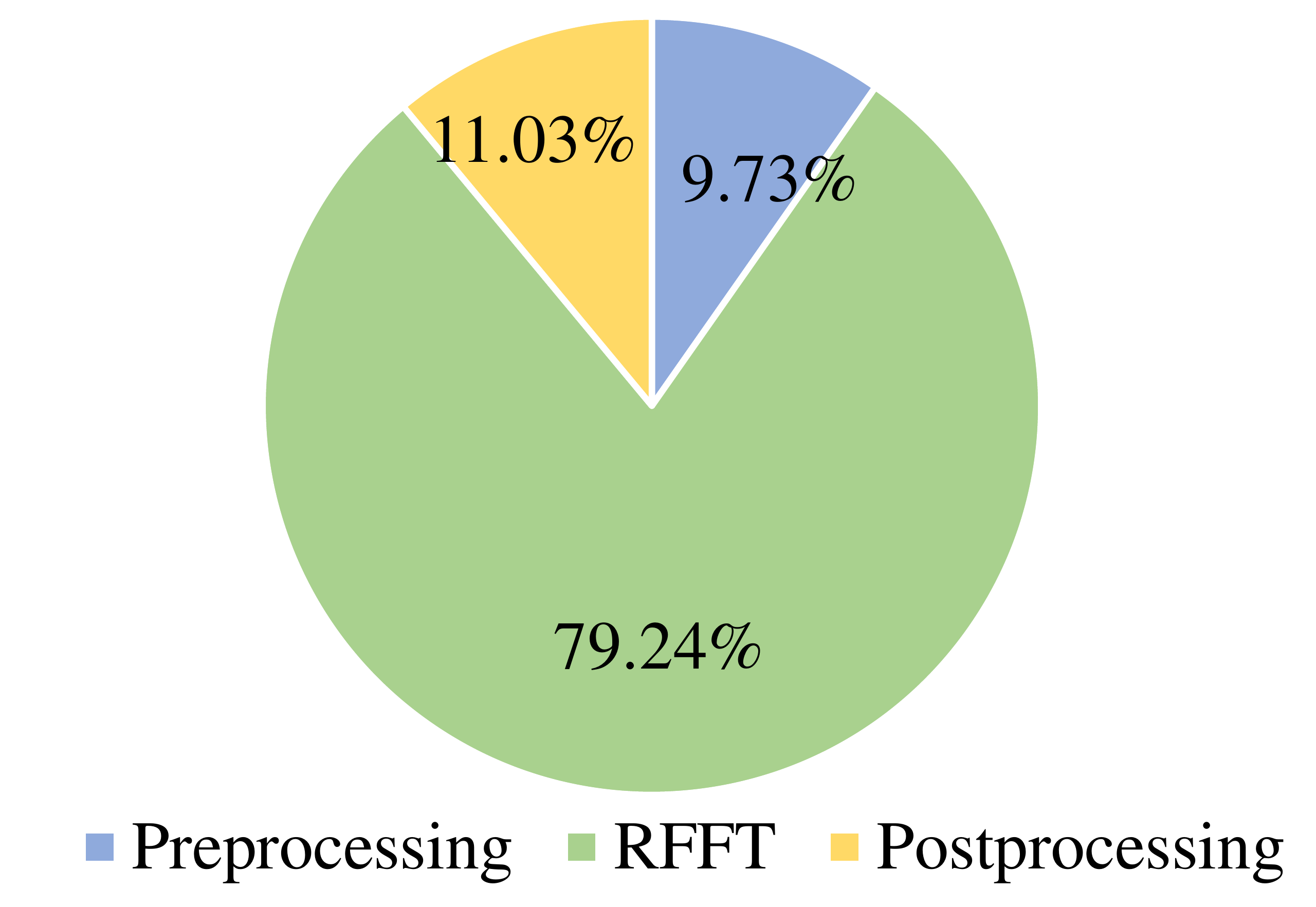}
    \caption{Runtime breakdown of our implemented 2D DCT ($N$=1024).}
    \label{fig:RuntimeBreakdown}
\end{figure}

\textbf{Execution time breakdown}.
We break down the execution time of our 2D DCT with an input size of $1024 \times 1024$ in Figure \ref{fig:RuntimeBreakdown}.
We observe that the RFFT takes most of the execution time, which means our implementation has comparable runtime to the highly-optimized FFT and is bottlenecked by the execution time of the RFFT.
Our preprocessing and postprocessing only take around 20\% of the execution time.
Postprocessing takes a relatively longer runtime than preprocessing due to extra numerical computation.
However, memory access in GPU global memory still bounds the postprocessing, especially when $N$ is large.
Thus the performance of those two stages already reaches the memory bound of the GPU roofline, which indicates that our implementation fully leverages the computation capability of GPUs.

\textbf{GPU utilization analysis}.
To evaluate the performance of our CUDA kernels, we use the NVIDIA visual profiler to report the GPU resource utilization in Table~\ref{tab:GPUUtilization}.
\begin{table}[t]
\centering
\caption{GPU utilization report for preprocessing and postprocessing CUDA kernels.
\textit{CTA} is cooperative thread array (CUDA block) and we use 256 threads per block. 
\textit{Occup.}, \textit{Comp.}, \textit{Mem. BW}, and \textit{SM Util.} represent GPU occupany, compute utilization, device memory bandwidth utilization, and streaming multiprocessor utilization, respectively.}
\label{tab:GPUUtilization}
\begin{tabular}{cccccc}
\hline\hline
{ Kernel}      & {CTA} & {Occup.} & { Comp.} & {Mem. BW} & { SM util.} \\\hline
{ preprocess}  & {16$\times$16}      & { 82.1\%}   & { 28.0\%} & { 78.1\%}      & { 100\%} \\
{ postprocess} & {16$\times$16}      & { 79.1\%}   & { 75.0\%} & { 75.6\%}      & { 99\%} \\\hline\hline  
\end{tabular}
\begin{tablenotes}
      \small
      \item \textit{CTA} represents cooperative thread array (CUDA block) and we use 256 threads per block. \textit{Occup.}, \textit{Comp.}, \textit{Mem. BW}, and \textit{SM Util.} represent GPU occupany, compute utilization, device memory bandwidth utilization, and streaming multiprocessor utilization, respectively.
\end{tablenotes}
\end{table}

For both kernels, we launch $16\times 16$ threads per cooperative thread array (CTA) for warp optimization.
Such a granularity for CUDA block can lead to near 100\% streaming multiprocessor (SM) utilization on all 30 SMs in a Titan Xp GPU, which means the workload balancing is nearly maximized for all threads.
Few double-precision computation resources are utilized such that our kernel performance is not bounded by computational resources.
As the primary task involved in our kernels is memory transaction, the device memory bandwidth is occupied by over 75\%.
The pre-computed exponential coefficients are assigned to the texture cache on GPUs, with higher bandwidth than global memory.
Therefore, the performance of our preprocessing and postprocessing kernels are all memory bounded according to the GPU roofline model.
With the memory coalescing technique being used, our CUDA kernels are already fully optimized on the target GPUs.

\section{Case studies}
\label{sec:case_studies}

DCTs are broadly used in scientific computing.
We provide two cases in this section: image compression and DREAMPlace~\cite{dreamplace}.
In the case study of DREAMPlace, we show how to apply our paradigm to other Fourier-related transforms.

We can estimate the speedup of an application when our proposed method is introduced.
Following Amdahl's law, the theoretical speedup of the execution of the whole task is shown in Equation~\eqref{eq:amdahl-law},
\begin{equation}
\label{eq:amdahl-law}
\frac{1}{(1-p) + p / s}
\end{equation}
where $p$ is the proportion of execution time that the Fourier-related transforms occupy in the original application,
$s$ is the speedup of our proposed method on the Fourier-related transforms.

\subsection{Image compression}

\begin{figure}[htbp]
    \centering
    \includegraphics[width=0.45\textwidth]{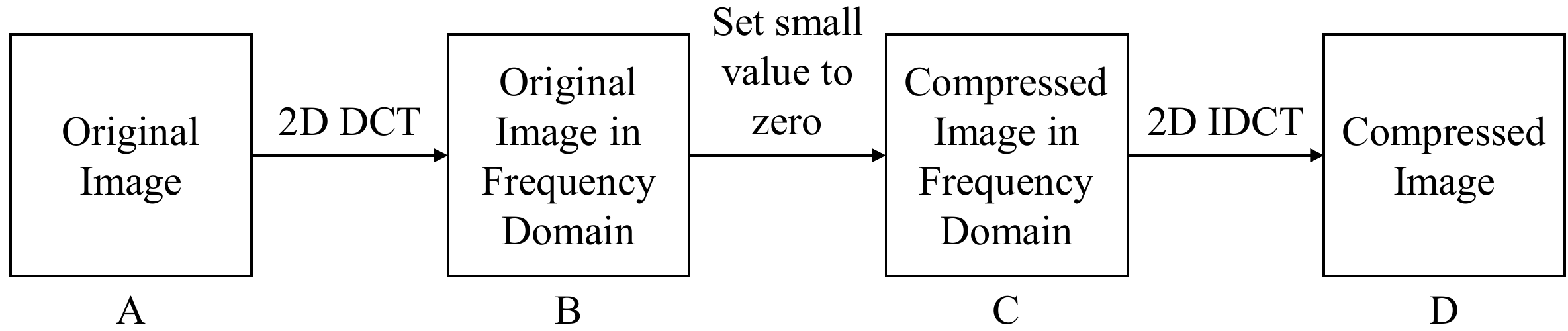}
    \caption{Image compression using a pair of 2D DCT and 2D IDCT.}
    \label{figure:image_compression}
\end{figure}

\begin{algorithm}[]
\begin{algorithmic}[1]
\caption{Image compression with 2D DCT and 2D IDCT}
\label{algorithm:image-compression}
\Require An image represented as the matrix $A$ in any shape
\State $B = \texttt{2D\_DCT}(A)$
\State $C = f_\epsilon(B)$ as shown in Equation~\eqref{equation:image-compression}
\State \Return $D = \texttt{2D\_IDCT}(C)$
\end{algorithmic}
\end{algorithm}

Figure~\ref{figure:image_compression} and Algorithm~\ref{algorithm:image-compression} illustrate a simple application of 2D DCT and 2D IDCT, which is the compression of 2D images.
We first translate the input image $A$ into the frequency domain to obtain $B$.
Then we can do magnitude-based compression in the frequency domain as shown in Equation~\eqref{equation:image-compression}.
\begin{equation}
    \label{equation:image-compression}
    C_{ij} = f_\epsilon(B_{ij}) = \begin{cases}
    B_{ij}, & \text{if } |B_{ij}| \geq \epsilon \\
    0, & \text{if } |B_{ij}| < \epsilon
    \end{cases}
\end{equation}
For each element $B_{ij}$, we set this entry to zero if its absolute value is smaller than a threshold $\epsilon$.
Finally, we translate the compressed image back to the time domain, with the result $D$.
The MATLAB documentation on dct2~\footnote{\url{https://www.mathworks.com/help/images/ref/dct2.html}} also uses this application as an example.
We validate our implementation against the MATLAB implementation.
If there are multiple channels in the input image, e.g., three channels in an RGB image, we can conduct compression in every channel.
This compression procedure, where DCT/IDCT is performed on the whole image, is different from the JPEG compression, where DCT/IDCT is performed on image blocks of $8 \times 8$ pixels.

Since we can fuse Line 2 of Algorithm~\ref{algorithm:image-compression} with the postprocessing of 2D DCT or the preprocessing of 2D IDCT,
we have $p = 1$ in the perspective of Amdahl's law and Equation~\eqref{eq:amdahl-law}.
Therefore, the compression will share the same speedup with our 2D DCT/IDCT implementations.

\subsection{DREAMPlace}
\begin{table*}[h]
\caption{Execution time of electric potential energy and electric force computation}
\label{table:dreamplace}
\centering
\begin{tabular}{ccccccccc}
\hline \hline
benchmark     & adaptec1 & adaptec2 & adaptec3 & adaptec4 & bigblue1 & bigblue2 & bigblue3 & bigblue4 \\ \hline
baseline (ms) & 0.63     & 2.00     & 2.27     & 2.88     & 0.67     & 2.56     & 8.25     & 15.10    \\
ours (ms)     & 0.33     & 1.00     & 1.29     & 1.89     & 0.38     & 1.53     & 4.87     & 11.72    \\
speedup  & 1.90     & 1.99     & 1.75     & 1.53     & 1.78     & 1.68     & 1.69     & 1.29     \\ \hline \hline
\end{tabular}
\begin{tablenotes}
\small
\item The baseline implementations use row-column methods, while ours implementation perform 2D transforms using 2D FFTs.
\end{tablenotes}
\end{table*}

Another example is using DCTs to solve Poisson's equation, which describes the potential field caused by a given charge or mass density distribution.
Given the potential field, we can then calculate the gravitational or electrostatic field and the final equilibrium.
Furthermore, we use our implementation in the DREAMPlace~\cite{dreamplace} framework, which needs to simulate an electrostatic system.

As the scale of very-large-scale integrated circuits (VLSI) grows, placement becomes a more challenging step for design closure since it is rather time-consuming. 
In placement problem, we try to assign locations for movable cells, such that the total half-perimeter wirelength is minimized.
There is a constraint that any two cells do not overlap.
DREAMPlace is a GPU-accelerated framework for the analytical VLSI placement. 
The layout and netlist are modeled as an electrostatic system, where standard cells in the circuits are modeled as electric charges.
Thus cells are driven by the electrostatic force to eliminate the overlaps~\cite{eplace}.
To get the numerical solution of electric potential distribution, the spectral method with Poisson's equation is adopted. 
When we use the spectral method, we usually need to define or combine several Fourier-related transforms due to the different boundary conditions.
Thus, highly-optimized 2D DCT/IDCT and other Fourier-related transforms are critical to achieving high performance.

In DREAMPlace, the authors define IDXST as follows,
\begin{equation}
\label{eq:idxst}
    \textrm{IDXST}(\{x_n\})_k = (-1)^k \textrm{IDCT}(\{x_{N-n}\})_k, \text{where } x_N = 0.
\end{equation}
They further define the following 2D transforms,
\begin{equation}
\begin{split}
    \textrm{IDCT\_IDXST}(x)&=\texttt{IDCT}(\texttt{IDXST}(x)^T)^T \\
    \textrm{IDXST\_IDCT}(x)&=\texttt{IDXST}(\texttt{IDCT}(x)^T)^T
\end{split}
\end{equation}
For IDCT\_IDXST, it can be calculated by first performing 1D IDCT along the row, then performing 1D IDXST along the column.
These two 2D transforms can also be computed through our paradigm, preprocessing, 2D IRFFT, and postprocessing.
The preprocessing and postprocessing are similar to those of 2D IDCT.
We validate our implementation against the row-column method.
The execution time of $\textrm{IDCT\_IDXST}$ for input sizes of $512 \times 512, 1024 \times 1024, 2048 \times 2048, 4096 \times 4096$ are $0.13$ ms, $0.42$ ms, $1.63$ ms, $6.80$ ms, similar to those of 2D IDCT.
Our standard procedure of preprocessing, FFT, and postprocessing can handle different Fourier-related transforms with rather stable performance.

\begin{algorithm}[]
\begin{algorithmic}[1]
\caption{Computation of electric potential and electric force in DREAMPlace}
\label{algorithm:electirc-dreamplace}
\State Compute the density map $\rho$
\State Compute electric potential $a = \texttt{2D\_DCT}(\rho)$
\State Compute scaled electric potential $a_1, a_2$ based on $a$
\State Compute electric force $\xi_1 = \texttt{IDCT\_IDXST}(a_1), \xi_2 = \texttt{IDXST\_IDCT}(a_2)$
\end{algorithmic}
\end{algorithm}

Algorithm~\ref{algorithm:electirc-dreamplace} is the simplified computation of electric potential and electric force in DREAMPlace.
2D DCT is used to calculate the electric potential, while the IDCT\_IDXST and IDXST\_IDCT are used to compute the electric force.
Then, the electrical charges are driven by the electric force to move.
We refer the readers to \cite{eplace, dreamplace} and the implementation \footnote{\url{https://github.com/limbo018/DREAMPlace}} for more details.

Table~\ref{table:dreamplace} compares the execution time of one step of the electric potential energy and electric force computations on ISPD 2005 benchmarks~\cite{ispd2005}.
We can achieve $1.7\times$ speedup on average.
The variance across different benchmarks can also be explained in the perspective of Amdahl's law and Equation~\eqref{eq:amdahl-law}.
In Table~\ref{table:dreamplace}, larger benchmarks, e.g., adaptec4 and bigblue4, have higher speedup on Fourier-related transforms (larger $s$ due to the larger input matrix) but considerably more extra computations in Lines 1 and 3 (smaller $p$) than smaller benchmarks, e.g., adaptec1 and bigblue1.
Hence, we observe smaller overall speedup on larger benchmarks.

\section{Conclusion}
\label{sec:Conclusion}

In this work, we propose a novel acceleration paradigm for MD Fourier-related transforms to achieve better parallelism, locality, and computation efficiency.
Our proposed general three-stage procedure can perform different transforms with a stable execution time regardless of transform types.
We use 2D DCT/IDCT CUDA implementation as an example to demonstrate the paradigm.
Our CUDA kernels eliminate computation redundancy via fully utilizing the conjugate symmetry in RFFT,
which leads to a higher arithmetic intensity and GPU utilization than prior methods.
The preprocessing and postprocessing procedures are highly optimized so that they introduce negligible overhead.
We will open-source our CUDA DCT implementations, which achieve $2\times$ speedup compared with the previous row-column methods.
Case studies demonstrate that our paradigm and implementations can achieve superior speedup in related applications.
We are convinced that our novel paradigm and high-performance CUDA implementation will boost the community of scientific computing,
especially in the field of signal processing and large-scale optimization.


\bibliographystyle{IEEEtranS}
\bibliography{reference}

\end{document}